\documentclass[iop,numberedappendix]{emulateapj}
\slugcomment{Accepted by \apj} 
\usepackage{natbib}
\usepackage{amsmath}
\usepackage{booktabs}
\usepackage{longtable}
\usepackage{color}
\usepackage{graphicx, subfigure}
\usepackage{hyperref}
\usepackage{CJKutf8}

\newcommand{\mr}{\mathrm} 
\newcommand{\di}{\mathrm{d}} 
\newcommand{\St}{\mathrm{St}} 

\newcommand{\rom}[1]{%
  \textup{\uppercase\expandafter{\romannumeral#1}}%
}

\defcitealias{OC2007}{OC2007}

\begin{document}
\title{Impact of magneto-rotational instability on grain growth in protoplanetary disks:
I. Relevant turbulence properties}
\author{Munan Gong (\begin{CJK*}{UTF8}{gbsn}龚慕南\end{CJK*})\altaffilmark{1}, Alexei V. Ivlev\altaffilmark{1}
Bo Zhao\altaffilmark{1}, and Paola Caselli\altaffilmark{1}}
\altaffiltext{1}{Max-Planck Institute for Extraterrestrial Physics,
Garching by Munich, 85748, Germany; munan@mpe.mpg.de}

\begin{abstract}
    Turbulence in the protoplanetary disks induces collisions between dust
    grains, and thus facilitates grain growth. We investigate the two fundamental
    assumptions of the turbulence in obtaining grain collisional velocities --
    the kinetic energy spectrum and the turbulence autocorrelation time -- in the
    context of the turbulence generated by the magneto-rotational instability
    (MRI). We carry out numerical simulations of the MRI as well as driven
    turbulence, for a range of physical and numerical parameters.
    We find that the convergence of the turbulence $\alpha$-parameter does not
    necessarily imply the convergence of the energy spectrum.
    The MRI turbulence is largely solenoidal, for which
    we observe a persistent kinetic energy spectrum of $k^{-4/3}$. The same is
    obtained for solenoidal driven turbulence with and without magnetic
    field, over more than 1 dex near the dissipation scale. This power-law
    slope appears to be converged in terms of numerical
    resolution, and to be due to the bottleneck effect.
    The kinetic energy in the MRI turbulence peaks 
    at the fastest growing mode of the MRI.
    In contrast, the magnetic energy peaks at the dissipation scale.
    The magnetic energy spectrum in the MRI turbulence 
    does not show a clear power-law range, and is almost constant over
    approximately 1 dex near the dissipation scale.
    The turbulence autocorrelation time is nearly constant at large scales,
    limited by the shearing timescale, and shows a power-law drop close to $k^{-1}$
    at small scales, with a slope steeper than that of the 
    eddy crossing time. The deviation 
    from the standard picture of the Kolmogorov turbulence with the injection scale
    at the disk scale height can potentially
    have a significant impact on the grain collisional velocities.
\end{abstract}

\section{Introduction}
Dust grains are the building blocks of planets. In order to build up
millimeter sized pebbles observed in protoplanetary disks,
and eventually planetesimals and planets, sub-micron sized interstellar
dust grains must coagulate and grow in size for many orders of magnitude. The
process of the dust grain growth in the protoplanetary disks 
has been modelled extensively in the literature \citep[see reviews
by][]{BM2008, Testi2014, Birnstiel2016}.
One of the fundamental parameter in
the grain growth models is the collisional velocity between dust grains. 
Turbulent motions of the gas stir up dust grains, and they are among the dominant
sources for collisional velocities of micro- to meter- sized grains 
\citep{Birnstiel2011}.

For the calculations of the grain collisional velocities induced by turbulence,
the Kolmogorov energy spectrum $E(k) \propto k^{-5/3}$ is the standard underlying 
assumption in the literature. In the ground-laying works of \citet{Voelk1980} and
\citet{Markiewicz1991}, the Kolmogorov spectrum is adopted. 
Subsequently an analytic solution is found by
\citet[][hereafter \citetalias{OC2007}]{OC2007},
which is widely used in many modern grain growth models 
\citep[e.g.][]{Brauer2008, Birnstiel2010, Garaud2013}. However, the Kolmogorov
spectrum describes the hydrodynamic (HD) turbulence, and 
protoplanetary disks are known to be magnetized. Many alternative theories have
been proposed for the magneto-hydrodynamic (MHD) turbulence. For example, the 
Iroshnikov–Kraichnan (IK) theory predicts
$E(k) \propto k^{-3/2}$ \citep{Iroshnikov1964, Kraichnan1965}, and the 
Goldreich–Sridhar theory predicts $E(k_\perp) \propto k_\perp^{-5/3}$
perpendicular to the mean magnetic field, and $E(k_\parallel) \propto
k_\parallel^{-2}$ parallel to the mean magnetic field. 
Moreover, in the case of a weak magnetic field, 
where the mean field orientation is varying on spacial and temporal scales of
the turbulence cascade, it is unclear whether the theoretical predictions by
the IK or Goldreich–Sridhar theories still hold.

One of the most likely mechanism for generating turbulence in the
protoplanetary disks is the MRI. Numerical simulations
also have been carried out to investigate the energy spectrum of turbulence
generated by the magneto-rotational instabilities (MRI) in the protoplanetary
disks \citep{Fromang2010, LL2011, Walker2016}. However, detailed studies of the
MRI turbulence and its consequence on the grain collisional velocities
in the protoplanetary disks have not been carried out before, 
and they are the aim of this work. Although non-ideal MHD effects are likely to be
important in the protoplanetary disks due to the low ionization fraction
\citep{Zhu2014, Bai2017, Simon2018}, we start with ideal MHD in this work, as a
natural first step before including more complex physics. Even though
the numerical dissipation is always present
and does not necessarily behave similar to the physical dissipation, the ideal
MHD simulation is
still a useful tool, where the effect of numerical dissipation can be studied
through varying the numerical resolution.
We also note that, while turbulence is widely assumed to be present in
protoplanetary disks, direct observational measurements 
have so far only been able to constrain the upper limits of the turbulent
velocities in the outer disk, due to the limitations in resolution and
sensitivity\citep{Flaherty2015, Flaherty2018, Teague2016, Pinte2016}.

In this paper, we investigate the energy spectrum and the autocorrelation time
of the MRI turbulence in the protoplanetary disks, for a range of physical and
numerical parameters. The structure of the paper is as follows: Section 
\ref{section:turbulence} introduces the turbulence properties that are
important for grain collisional velocities, which we focus on in the rest of
the paper. Section \ref{section:method} describes the numerical methods, and
Section \ref{section:results} states the results. Finally, Section
\ref{section:conclusions} summaries the conclusions.

For the convenience of the reader, the notations for important physical variables
are summarized in Table \ref{table:notations}.
\begin{table}[htbp]
    \centering
    \caption{Summary of notations for the key physical variables}
    \label{table:notations}
    \begin{tabular}{cc}
        \tableline
        \tableline
        Symbol &Meaning\\
        \tableline
        $\rho_0$ &code unit for density\\
        $\Omega_0$ &code unit for frequency\\
        $L_0$ &code unit for length\\
        $t_0$ &code unit for time\\
        $t_\mr{orbit}$ &$2\pi/\Omega_0$, orbital period\\
        $c_s$ &sound speed, $c_s=1$ in code unit\\
        $\Omega$ &local Keplerian orbital frequency\\
        $H$      &$c_s/\Omega$, disk scale-height\\
        $\beta$  &$2c_s^2\rho/B^2$, plasma beta\\
        $\rho_\mr{init}$ &initial (mid-plane) density\\
        $B_\mr{init}$    &initial vertical magnetic field strength\\
        $\beta_\mr{init}$
        &$2c_s^2\rho_\mr{init}/B_\mr{init}^2$, initial (mid-plane) plasma beta\\
        $Q_z$    &quality parameter for resolution (Eq. (\ref{eq:Qz}))\\
        $\mathbf{v}_K$  &Keplerian velocity\\
        $\mathbf{v}$    &gas velocity\\
        $\delta \mathbf{v}$ & $\mathbf{v} - \mathbf{v}_K$, turbulent gas velocity\\
        $v_\mathrm{tot}$ &total gas turbulent velocity (Eq. (\ref{eq:vt}))\\
        $\tau(k)$ &eddy auto-correlation time (Eq. (\ref{eq:tauk})) \\
        $P(k)$ &turbulence kinetic power spectrum (Eq. (\ref{eq:Pk}))\\
        $E(k)$ &$4\pi k^2 P(k)$, kinetic energy spectrum\\ 
        $P_M(k)$ &magnetic power spectrum (Eq. (\ref{eq:PM}))\\
        $E_M(k)$ &$4\pi k^2 P_M(k)$, magnetic energy spectrum\\ 
        $\alpha_R$ &Reynolds stress (Eq. (\ref{eq:alphaK}))\\ 
        $\alpha_\mr{Maxw}$ &Maxwell stress (Eq. (\ref{eq:alphaM}))\\ 
        $\alpha$ &total turbulence stress (Eq. (\ref{eq:alpha}))\\ 
        $p$ &slope in power-law range $E(k)\propto k^{-p}$\\
        $p_M$ &slope in power-law range $E_M(k)\propto k^{-p_M}$\\
        $m$ &slope in power-law range $\tau(k)\propto k^{-m}$\\
        $\widetilde{y}$ &Fourier transform of variable $y$ (Eq. (\ref{eq:ft}))\\
        $\langle y \rangle$ &spacial average of variable $y$
                             (text below Eq. (\ref{eq:alphaM}))\\
        $\overline{y}$ &time average of variable $y$ during steady state\\
        \tableline
        \tableline
    \end{tabular}
\end{table}

\section{Relevant Turbulence Properties}\label{section:turbulence}
The collisional velocities between dust grains induced by turbulent motions can
be calculated based on the (semi-)analytic framework introduced by \citet{Voelk1980},
\citet{Markiewicz1991} and \citetalias{OC2007}. In this framework, 
the collisional velocity of two dust grains of certain sizes
is determined by two important properties of the turbulence: 
the kinetic power spectrum $P(k)$, and the auto-correlation time $\tau(k)$.
We define these quantities below. 

The turbulence kinetic power spectrum
$P(k)$ is defined as $P(k) = P_x(k) + P_y(k) + P_z(k)$, where 
\begin{equation}\label{eq:Pk}
    P_j(\mathbf{k}) = \frac{1}{L_x L_y L_z}|\widetilde{\delta v}_j(\mathbf{k})|^2,
    ~j=x, y, z.
\end{equation}
$\widetilde{\delta v}_j(\mathbf{k})$ is the Fourier transform of the component
$\delta v_j(\mathbf{x})$ of the turbulent velocity field.\footnote{To obtain the turbulent
velocity field in the shearing-box simulations, 
the systematic velocity from Keplerian shear is subtracted.}
$P_j(k)$ is the average of $P_j(\mathbf{k})$ at a constant
magnitude of $k=|\mathbf{k}|$. The energy spectrum is defined as\footnote{
    Note that the $E(k)$ defined in
\citetalias{OC2007} is a factor of two smaller than our definition.}
$E(k) \equiv 4\pi k^2 P(k)$.
In this paper, the 3-dimensional Fourier transform is defined as
\begin{equation}\label{eq:ft}
    \widetilde{y}(\mathbf{k}) = \frac{1}{(2\pi)^{3/2}} 
       \int y(\mathbf{x}) \exp(-i \mathbf{k}\cdot\mathbf{x}) \di^3 \mathbf{x}.
\end{equation}
The total turbulent velocity $v_\mathrm{tot}$ is defined as
\begin{equation}\label{eq:vt}
    v_\mathrm{tot}^2 = \int \di k E(k).
\end{equation}
From the Plancherel theorem, $v_\mathrm{tot}^2 = \langle \delta v^2 \rangle$.

To understand the property of turbulence, it is helpful to decompose the
turbulent velocity field into compressive and solenoidal modes. The
decomposition is straight-forward in the Fourier space, where 
$\widetilde{\delta \mathbf{v}}=\widetilde{\delta \mathbf{v}}_s +
\widetilde{\delta \mathbf{v}}_c$; here, $\widetilde{\delta \mathbf{v}}_c =
(\widetilde{\delta \mathbf{v}} \cdot \mathbf{k})\mathbf{e}_k$ is the compressive
mode, where $\mathbf{e}_k$ is the unit vector in the direction of $\mathbf{k}$.
For isotropic turbulence, $\langle \delta v_s^2 \rangle = 2\langle \delta v_c^2
\rangle = 2\langle \delta v^2 \rangle/3$.

We also define the magnetic power spectrum
\begin{equation}\label{eq:PM}
    P_M(\mathbf{k}) = \frac{1}{L_x L_y L_z}|\widetilde{B}(\mathbf{k})|^2,
\end{equation}
and the magnetic energy spectrum $E_M(k) \equiv 4\pi k^2 P_M(k)$.

The turbulence auto-correlation time $\tau(k)$
is defined following \citet{Markiewicz1991} (their Equation (5)),
\begin{equation}\label{eq:tauk}
\begin{split}
    \widetilde{\delta \mathbf{v}}(\mathbf{k},t)\cdot
      \widetilde{\delta \mathbf{v}}^*(\mathbf{k},t')
    = &|\widetilde{\delta \mathbf{v}}(\mathbf{k},t)|^2\\
        &~\times
        \left( 1 + \frac{|t-t'|}{\tau(k)}\right)\exp\left(
        - \frac{|t-t'|}{\tau(k)}\right).
\end{split}
\end{equation}

Grain collisional velocities in turbulent gas depend on the exact shape of
$E(k)$ and $\tau(k)$, especially for small grains that are well-coupled to the
gas. The dynamic properties of the dust grain can be characterized by the 
dimensionless Stokes number $\St = \tau_f \Omega$, where $\tau_f$ is the
friction/stopping time of the dust grain. For two small dust grains both with
$\St < 1$, the collisional velocity is dominated by the turbulence eddy 
that the {\sl larger} grain first decouples from. If the larger grain has a
friction time $\tau_{f}$ and Stokes number $\St$, the eddy $k$ with 
$\tau(k)\approx \tau_{f}$ contributes the most to the collisional velocity.
For $E(k)\propto k^{-p}$ and $\tau(k) \propto k^{-m}$, the
collisional velocity 
\begin{equation}
v^2_\mr{coll}\propto k^* E(k^*) \propto \St^{(p-1)/m}.
\end{equation}
Assuming $\tau(k) = 1/(k\sqrt{2kE(k)})$, 
for Kolmogorov turbulence, we have $p=5/3$ and
$m=2/3$ with $v^2_\mr{coll}\propto \St$; for IK turbulence, we have
$p=3/2$ and $m=3/4$ with $v^2_\mr{coll}\propto \St^{2/3}$.
If we take the values $p=4/3$ and $m=1$ observed
in this study,\footnote{In Section
\ref{section:results}, we show
that the values of $p$ and $m$ observed in this work may not necessarily
represent the values in the true inertial range and be extrapolated to
large scales. However, we still use this example here to show that the
collisional velocity between small grains is sensitive to the exact shape of
$E(k)$ and $\tau(k)$.} then $v^2_\mr{coll}\propto \St^{1/3}$.  

Consider spherical dust particles in the Epstein drag regime \citep{Epstein1924}, 
the friction time $\tau_f = \rho_p a/(\rho c_s)$, where $\rho_p$ and $a$ are
the density and radius of the dust particle. Using the minimum mass solar
nebular model (MMSN) in \citet{Hayashi1981}, the Stokes number can be written as
$\St = \rho_p a\Omega/(\rho c_s) = 1.4\times 10^{-4} (a/\mathrm{mm})
(R/\mathrm{AU})^{4.5}$, where $R$ is the radial location in the disk.
For a $1~\mathrm{\mu m}$ dust grain located at
$1~\mr{AU}$ radius in the protoplanetary disk, the Stokes number can be as
small as $10^{-7}$, and {\sl thus a slight variation in the shapes of
$E(k)$ and $\tau(k)$
can change the collisional velocities by orders of magnitude.} Therefore, it is
important to understand the behavior of $E(k)$ and $\tau(k)$ in order 
to constrain the grain growth models.

In this paper, we focus on studying
$E(k)$ and $\tau(k)$ from the MRI turbulence. We defer the detailed investigation
of the dependence of grain collisional velocities on $E(k)$ and $\tau(k)$ to a
following paper.

\section{Numerical Method}\label{section:method}
We perform two sets of simulations. In order to investigate the properties of
MRI turbulence in the protoplanetary disk, we perform ideal MHD local shearing-box
simulations with a net vertical magnetic field.\footnote{We include a net
vertical magnetic field because it is likely to be present in the disk,
and simulations with zero net vertical magnetic flux show issues of
numerical convergence \citep{Shi2016}.}
For comparison, we also carry
out driven turbulence simulations with and without magnetic fields, where
kinetic energy is continuously injected at each simulation timestep 
at large scales. The details of both sets of simulations are described below.

\subsection{MRI Simulations\label{section:method:MRI}}
Our MRI simulations are conducted using the {\sl
Athena} code \citep{Stone2008, SG2009}. We perform three-dimensional ideal 
MHD simulations with the local shearing-box approximation \citep{SG2010}, in a
reference frame corotating with the disk at Keplerian
orbital frequency $\Omega$ at a fiducial radius. 
We adopt a Cartesian coordinate system $(x, y, z)$, with $\mathbf{e}_x$,
$\mathbf{e}_y$ and $\mathbf{e}_z$ denoting respectively the unit vectors in
radial, azimuthal, and vertical directions.

The ideal MHD equations reads\footnote{In Equations (\ref{eq:mass})--(\ref{eq:energy}),
the $1/\sqrt{4\pi}$ pre-factor is absorbed in the unit of $\mathbf{B}$.}
\begin{equation}\label{eq:mass}
    \frac{\partial \rho}{\partial t} + \nabla\cdot(\rho \mathbf{v}) = 0,
\end{equation}
\begin{equation}\label{eq:momentum}
    \begin{split}
        &\frac{\partial \rho\mathbf{v}}{\partial t} 
    + \nabla\cdot(\rho \mathbf{vv} - \mathbf{BB})
    + \nabla\left( P + \frac{1}{2}B^2\right)\\
        &= 2\rho q \Omega^2 \mathbf{x} - \rho \Omega^2 \mathbf{z} 
      - 2 \Omega \mathbf{e}_z \times \rho \mathbf{v},
    \end{split}
\end{equation}
\begin{equation}\label{eq:energy}
    \frac{\partial \mathbf{B}}{\partial t} 
    - \nabla \times (\mathbf{v} \times \mathbf{B}) = 0,
\end{equation}
where $\rho$ is the mass density, $\mathbf{v}$ is the gas velocity, $P$ is the gas
pressure, and $\mathbf{B}$ is the magnetic field. The shear $q$ is defined as
\begin{equation}
    q = - \frac{\di \ln \Omega}{\di \ln r},
\end{equation}
and we use $q=3/2$ for Keplerian disks. We assume an isothermal equation of
state $P=c_s^2\rho$, where $c_s$ is the isothermal sound speed.  
The term $- \rho \Omega^2 \mathbf{z}$ on the right-hand side of the momentum
Equation (\ref{eq:momentum}) is the vertical gravity of the disk. For most of
the simulations in this paper, we focus on the mid-plane of the disk and ignore
the vertical gravity (unstratified shearing-box). For comparison, 
we also perform one simulation including the vertical gravity
(stratified shearing-box, see Table \ref{table:simulations}).

For all the simulations, the HLLD Riemann solver \citep{MK2005} is used, and 
an orbital advection scheme \citep[FARGO, see][]{Masset2000, Johnson2008} is
adopted to reduce the truncation error induced by the background shear. For the
unstratified simulations, we use the Corner Transport Upwind (CTU) integrator
and third-order spacial reconstruction. For the stratified simulation, we use
the van Leer (MUSCL–Hancock type) integrator and second-order spacial
reconstruction to improve the stability of the code. The radial ($x$) boundary
condition is shearing periodic, the azimuthal ($y$) boundary condition is
periodic, and the vertical ($z$) boundary condition is periodic for
unstratified simulations and outflow for the stratified simulation.
The simulations are performed on a uniform Cartesian grid. The grid cells are
cubic with length $\Delta x$ in each dimension. The simulation domain has a
box-size of $L_x \times L_y \times L_z$.

We use the density, velocity, and orbital frequency units of 
$\rho_0=v_0=\Omega_0=1$ in the code, and adopt $c_s=1$ in all simulations.
The initial density of the stratified simulation is set to be the equilibrium
Gaussian density profile $\rho=\rho_\mr{init} \exp(-z^2/2H^2)$, 
where the mid-plane gas density $\rho_\mr{init} = \rho_0 = 1$, and the 
disk scale-height $H=c_s/\Omega$. 
For the unstratified simulations, we use an initial constant
density of $\rho_\mr{init}=0.747\rho_0$, same as the average density within $|z|<\sqrt{2}H$
in the stratified simulation. We set the initial velocity to be the Keplerian
orbital motion 
\begin{equation}\label{eq:vk}
\mathbf{v}_K=-q\Omega x \mathbf{e}_y.
\end{equation}
We put small initial random perturbations on the 
density and velocity fields. We set an initial uniform vertical magnetic field
$\mathbf{B}_\mr{init}$. We also add an additional zero-net-flux sinusoidal field 
$\mathbf{B}_\mr{init}' = B_\mr{init} \sin (2\pi x/L_x) \mathbf{e}_z$ 
to make the simulation more stable at early times;
this does not affect the steady state of the simulation later on. The
magnetic field strength is characterized by the plasma
$\beta=2c_s^2\rho/B^2$. The initial vertical magnetic field has 
$\beta_\mr{init}=2 c_s^2 \rho_\mr{init}/B_\mr{init}^2=6.25\times 10^3$. We vary other numerical and
physical parameters of the simulations in our different models, which are
summarized in Table \ref{table:simulations}. The naming convention of the
models is ``B[log of the initial plasma $\beta$]L[box-size]R[resolution]''. 
The box-size and resolution is
given in the unit of $L_0 = c_s/\Omega_0$, and $L_0 = 1$ in code units. For
model B4L2R64-O3 with faster Keplerian rotation,
the disk scale-height $H=c_s/(3\Omega_0)=L_0/3$, and for all
other simulations $H=c_s/\Omega_0=L_0$. Because the stratified model
B5L2R32-S requires a larger box and has shorter time steps due to high-speed
wind near the vertical boundary, it is computationally more expensive. 
Thus, it is run for a shorter time of $80t_\mr{orbit}$, where $t_\mr{orbit}=2\pi/\Omega_0$.
All other simulations are run for $130t_\mr{orbit}$.

To understand the overall properties of the MRI turbulence and compare with
the existing literature, we examine the non-dimensional
Reynolds stress $\alpha_R$ and Maxwell stress $\alpha_\mr{Maxw}$ responsible for the
angular momentum transport,
\begin{equation}\label{eq:alphaK}
    \alpha_R = \frac{\langle \rho v_x \delta v_y \rangle}{\langle \rho c_s^2
    \rangle},
\end{equation}
and
\begin{equation}\label{eq:alphaM}
    \alpha_\mr{Maxw} = \frac{\langle -B_x B_y \rangle}{\langle \rho c_s^2 \rangle},
\end{equation}
where the angle brackets denote the spacial average over the
whole simulation domain for unstratified simulations, and over the mid-plane region
($|z|<\sqrt{2}H$) for the stratified simulation. 
The total turbulence $\alpha$-parameter is
\begin{equation}\label{eq:alpha}
    \alpha = \alpha_R + \alpha_\mr{Maxw}.
\end{equation}

To ensure the turbulence is well resolved, we compute the quality parameter in
the vertical direction $Q_z$ \citep{Hawley2011}:
\begin{equation}\label{eq:Qz}
    Q_z = \frac{2\pi \overline{\langle v_{A,z} \rangle}}{\Omega \Delta x}, 
\end{equation}
where the $\langle v_{A,z} \rangle = \sqrt{\langle B_z^2 \rangle/\langle \rho
\rangle}$, and the bar denotes the time average in steady state.
\citet{Sano2004} found that $Q_z \gtrsim 6$ is required for the turbulence
$\alpha$-parameter to be converged in the MRI steady state.

\begin{table*}[htbp]
    \centering
    \caption{Parameters for the MRI simulations}
    \label{table:simulations}
    \begin{tabular}{cccc ccc}
        \tableline
        \tableline
        Model ID & box-size $\frac{L_x}{L_0} \times \frac{L_y}{L_0} \times
        \frac{L_z}{L_0}$ &resolution per $L_0$
        &$\beta_\mr{init}=2 c_s^2 \rho_c/B_\mr{init}^2$ &$\Omega/\Omega_0$ 
        &stratified?  &Duration ($t_\mr{orbit}$)\\
        \tableline
        B5L2R32 &$2\times 2 \times 2$ &32 &$10^5$ &1   &no   &130\\
        B5L2R64 &$2\times 2 \times 2$ &64 &$10^5$ &1   &no   &130\\
        B5L2R128 &$2\times 2 \times 2$ &128 &$10^5$ &1   &no   &130\\
        B5L1R128 &$1\times 1 \times 1$ &128 &$10^5$ &1   &no   &130\\
        B5L1R256 &$1\times 1 \times 1$ &256 &$10^5$ &1   &no   &130\\
        B5L4R64 &$4\times 4 \times 2$ &64 &$10^5$ &1   &no   &130\\
        B4L2R64 &$2\times 2 \times 2$ &64 &$10^4$ &1   &no   &130\\
        B4L2R64-O3 &$2\times 2 \times 2$ &64 &$10^4$ &3   &no   &130\\
        B5L2R32-S &$2\times 2 \times 16$ &32 &$10^5$ &1   &yes   &80\\
        \tableline
        \tableline
    \end{tabular}
\end{table*}

\subsection{Driven Turbulence Simulations}
We perform the driven turbulence simulations using the code {\sl Athena++}
\citep{White2016, Stone2019}, a recent redesign of the code {\sl
Athena}.\footnote{We used {\sl Athena} for MRI simulations because the orbital
advection scheme is not yet implemented in {\sl Athena++}. Moreover, time
correlated driving of the turbulence is implemented in {\sl Athena++} but not
yet in {\sl} Athena. As shown by \citet{Grete2018}, uncorrelated turbulence
driving can inject additional compressive modes and change the energy spectrum
of the turbulence.} The equations
solved are similar to those in the MRI simulations in Section
\ref{section:method:MRI}, but without the terms from Keplerian shear and
vertical gravity (the right hand side of Equation (\ref{eq:momentum}) is zero).
Similar to the MRI simulations, we adopt a Cartesian coordinate system and the
isothermal equation of state. The boundary conditions are periodic on all
sides.

We carry out both hydrodynamic (HD) and ideal MHD simulations
with an initial net vertical magnetic field.  The HD and MHD simulations
are performed using the ROE and HLLD Riemann solvers respectively, and with
the third-order Runge-Kutta integrator and third order spacial
reconstruction in both cases.
The code units are the same as in the MRI simulations (Section
\ref{section:method:MRI}). We use a cubic box with $L_x = L_y = L_z = L_0$, and
the grid cells are also cubic.

The initial condition is set with a uniform density field $\rho_\mr{init} = 1$,
velocity field $\mathbf{v}=0$, and $c_s=1$. The cases with magnetic field has
an initial $\beta_\mr{init}=10^5$. The turbulence is continuously
driven with a stochastic forcing method described by \citet{Schmidt2009}. At
each computational time step, an additional forcing velocity field is added to
the simulation domain at large scales $0 < |\mathbf{k}| < 2$, with an energy
spectrum\footnote{The final energy spectrum in the
power-law range far away from the injection scales does not depend on the
energy spectrum of the turbulence injection.} $E(k) \propto k^{-3}$.
The forcing velocity field is either isotropic
or fully compressive. We also tested the case of fully solenoidal driving, and
obtained results similar to the isotropic driving case, where the solenoidal mode
still dominates. We inject kinetic energy with a constant $\di E/\di t = 0.001$
in code units, and set the forcing correlation time to $T_\mathrm{corr}=0.5 t_0$.
Here $T_\mathrm{corr}$ is large enough, so that the turbulence energy
spectrum and correlation time in the steady state are not affected by
$T_\mathrm{corr}$ \citep{Grete2018}. We run the simulations for $90 t_0$, about
9 times the turbulent crossing time. 
The simulation parameters for different models are
summarized in Table \ref{table:simulations_dr}.
\begin{table}[htbp]
    \centering
    \caption{Parameters for the driven turbulence simulations}
    \label{table:simulations_dr}
    \begin{tabular}{cccc}
        \tableline
        \tableline
        Model ID  &resolution per $L_0$ &driving &B-field?\\
        \tableline
        DR200     &200        &isotropic     &no\\
        DR400     &400        &isotropic     &no\\
        DR200C    &200        &compressive   &no\\
        DR400C    &400        &compressive   &no\\
        DR200B5   &200        &isotropic     &yes\\
        DR200B5C  &200        &compressive   &yes\\
        \tableline
        \tableline
    \end{tabular}
\end{table}

\section{Results}\label{section:results}
\subsection{Global Properties\label{section:results:MRI}}
\subsubsection{MRI Simulations}
\begin{figure}[htbp]
\centering
\includegraphics[width=\linewidth]{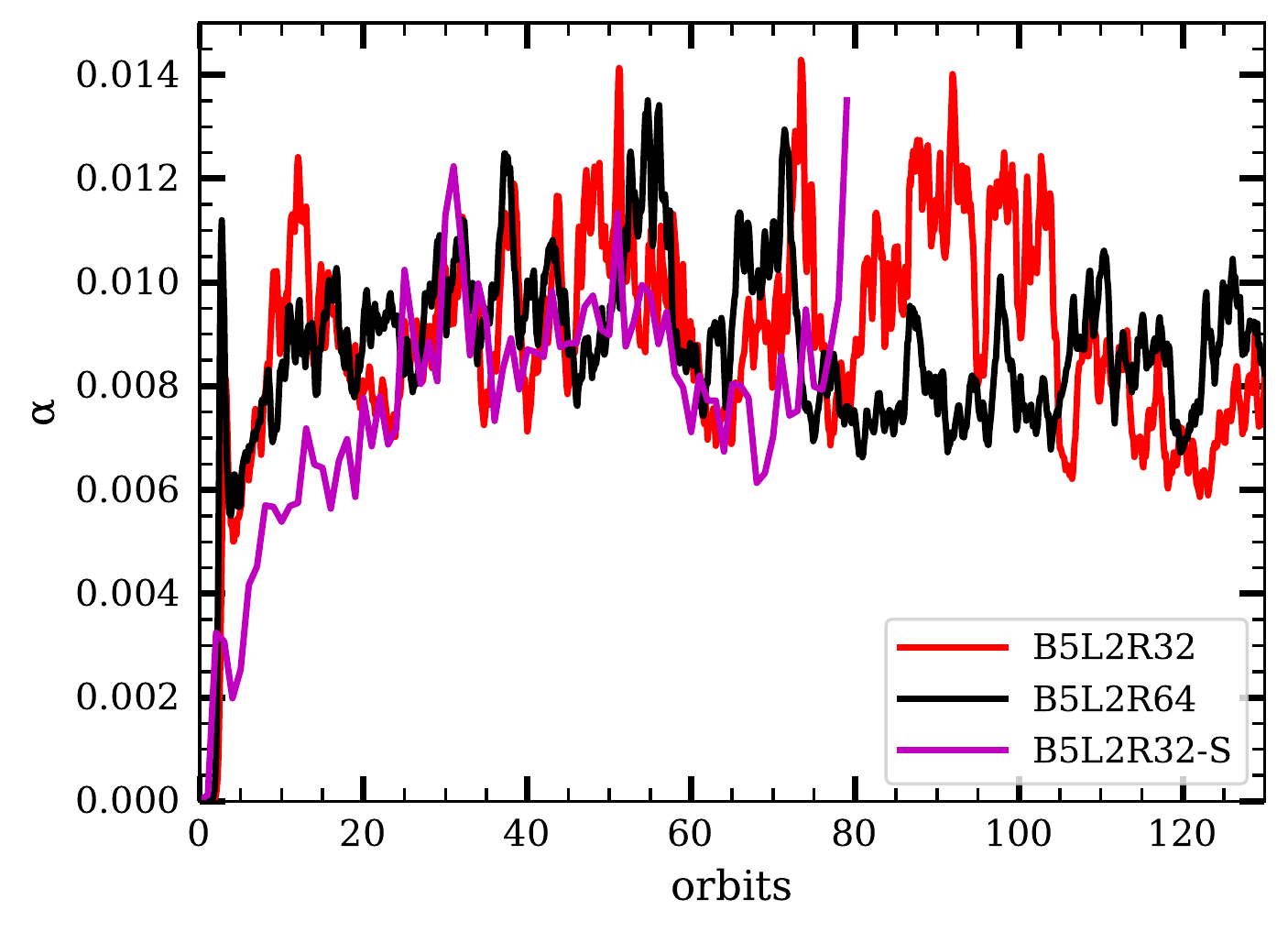}
\caption{The evolution of the $\alpha$-parameter. $\alpha$ increases
    initially and reaches the steady state after about 10 orbits for the unstratified
    simulations (B5L2R32 and B5L2R64) and after about 30 orbits for the
    stratified simulation (B5L2R32-S). The unstratified
    simulation B5L2R32 and stratified simulation B5L2R32-S give very similar
    values of $\alpha$ in steady states (see also Table
    \ref{table:global}). $\alpha$ is already converged with a
    numerical resolution of 32 cells per $L_0$ (B5L2R32), although the
    kinetic energy spectrum is not converged until a much higher resolution is
    used (see Figure \ref{fig:Pk_MRI} and discussions).
    Similar behaviors of $\alpha$ are found in all other simulations
    that are not plotted here.}
\label{fig:alpha}
\end{figure}

Initially, the MRI develops exponentially in the linear regime. After about
10--30 orbits, the MRI saturates in the non-linear regime, and the simulation
reaches a steady state. This general behavior is found in many similar previous
numerical studies of the MRI \citep[e.g.][]{Hawley1995, Stone1996, Simon2012,
BS2013}. To show the development and saturation of the MRI, we plot the 
evolution of the $\alpha$-parameter in Figure \ref{fig:alpha}. In all of our
simulations, $\alpha$ increases initially, until it reaches a steady state, and
oscillates around a constant value. Similar behaviors are found in other global
diagnostics of the turbulence, such as the kinetic and magnetic energies. 

The numerical effect of resolution and box-size can be seen from the global
properties of the different models in Table \ref{table:global}. For the models
with relatively large box-sizes, $L_x, L_y, L_z \geq 2L_0$, the global
properties such as the $\alpha$-parameter
are converged with a resolution of 32 cells per
$L_0$ (B5L2R32, B5L2R64 and B5L2R128). For models with smaller box-sizes,
$\alpha$-parameter and other quantities show anomalous behavior, changing with both
box-size and resolution. Similar phenomenon is found in \citet{Simon2012}.
$\alpha$-parameter increases with both the initial magnetic field strength 
(B4L2R64) and the rotational frequency (B4L2R64-O3). The global mid-plane 
properties in
the stratified model B5L2R32-S are very similar to the unstratified model
B5L2R32, justifying our approximation of ignoring the vertical gravity in the
rest of the models.

The convergence of $\alpha$-parameter, however, does not necessarily 
imply the convergence of turbulence kinetic energy spectrum $E(k)$.
In Section \ref{section:result:Pk}, 
we show that a much higher resolution of at least 128
cells per $L_0$ is needed in order to resolve the power-law range of
$E(k)$ near the dissipation scale. 

\begin{table*}[htbp]
    \centering
    \caption{Global properties of the MRI simulations 
             in steady states\tablenotemark{a}}
    \label{table:global}
    \begin{tabular}{cccc ccc c}
        \tableline
        \tableline
        Model  &$\sqrt{\langle \delta v^2 \rangle}/c_s$
        &$\sqrt{\langle v_{A,z}^2 \rangle}/c_s$
        &$\sqrt{\langle B_z^2 \rangle}/B_\mr{init}$
        &$\alpha_R/10^{-3}$ &$\alpha_\mr{Maxw}/10^{-3}$ &$\alpha/10^{-3}$ &$Q_z$\\
        \tableline
B5L2R32    &$0.14\pm 0.02$ &$0.20\pm 0.02$ &$8.63\pm 1.33$ &$1.68\pm 0.46$ &$7.06\pm 1.61$
           &$8.74\pm 2.02$ &$8$\\
B5L2R64    &$0.13\pm 0.01$ &$0.20\pm 0.01$ &$9.50\pm 0.62$ &$1.49\pm 0.19$ &$6.85\pm 0.78$
           &$8.35\pm 0.94$ &$17$\\
B5L2R128   &$0.14\pm 0.01$ &$0.22 \pm 0.02$ &$11.5\pm 0.7$  &$1.49\pm 0.18$ &$7.89\pm 0.82$
           &$9.38\pm 0.95$ &$41$\\
B5L1R128   &$0.10\pm 0.01$ &$0.16 \pm 0.02$ &$9.52\pm 1.00$ &$0.83\pm 0.18$ &$4.28\pm 0.79$
           &$5.12\pm 0.93$ &$34$\\
B5L1R256   &$0.14\pm 0.01$ &$0.23 \pm 0.03$ &$15.1\pm 1.6$  &$1.29\pm 0.36$ &$8.03\pm 1.39$
           &$9.32\pm 1.66$ &$109$\\
B5L4R64    &$0.15\pm 0.01$ &$0.22 \pm 0.01$ &$10.4\pm 0.6$  &$2.02\pm 0.25$ &$8.20\pm 0.86$
           &$10.2\pm 1.0$  &$19$\\
B4L2R64    &$0.24\pm 0.02$ &$0.40 \pm 0.05$ &$7.71\pm 0.82$ &$4.78\pm 1.09$ &$23.6\pm 3.6$ 
           &$28.4\pm 4.3$  &$44$\\
B4L2R64-O3 &$0.30\pm 0.03$ &$0.46 \pm 0.04$ &$7.53\pm 0.86$ &$9.04\pm 1.46$ &$33.6\pm 5.3$
           &$42.6\pm 6.7$  &$14$\\
B5L2R32-S  &$0.15\pm 0.01$ &$0.26 \pm 0.09$ &$8.69\pm 0.82$ &$1.69\pm 0.28$ &$6.93\pm 1.10$  
           &$8.62\pm 1.30$ &$8$\\
        \tableline
        \tableline
    \end{tabular}
    \tablenotetext{1}{The average value followed by the standard deviation
    (only the average is shown for $Q_z$), over simulation time $40-80t_\mr{orbit}$
    for model B5L2R32-S and $90-130t_\mr{orbit}$ for
    all other models. Note that the quantities in the stratified model
    B5L2R32-S are taken from the mid-plane region $|z|<\sqrt{2}H$. }
\end{table*}

\subsubsection{Driven Turbulence Simulations}
As the energy is injected into the simulation domain,
the kinetic (and magnetic, for MHD simulations) energy increases initially,
and reaches a steady state after $t\gtrsim 40 t_0$,
about 4 times the turbulent crossing time. We run the simulations
until $90 t_0$, and analyse the results using the outputs in the steady state
during $50-90 t_0$. 

The global properties of the simulations in steady states are summarized in
Table \ref{table:global_dr}. The average turbulent velocity, 
$\sqrt{\langle \delta v^2 \rangle}/c_s \approx 0.10-0.15$,
is similar to that in the MRI simulations (Table \ref{table:global_dr}). In the MHD
simulations, the magnetic energy is similar to the
kinetic energy in the isotropic driving case (DR200B5), but is only a small
fraction of the kinetic energy in the compressive driving case (DR200B5C). For
the HD simulations, the total kinetic energy is converged with a
resolution of 200 grid cells per $L_0$, for both the isotropic and compressive
driving. Due to the constraints on computational resources, the MHD simulations
are only performed at a resolution of 200 grid cells per $L_0$.
\begin{table}[htbp]
    \centering
    \caption{Global properties of the driven turbulence simulations in steady
    states\tablenotemark{a}}
    \label{table:global_dr}
    \begin{tabular}{c cc}
        \tableline
        \tableline
        Model ID  &$\sqrt{\langle \delta v^2 \rangle}/c_s$
        &$\sqrt{\langle v_{A,z}^2 \rangle}/c_s$ \\
        \tableline
        DR200     &$0.14\pm 0.003$ &-\\
        DR400     &$0.15\pm 0.002$ &-\\
        DR200C    &$0.10\pm 0.01$  &-\\
        DR400C    &$0.10\pm 0.01$  &-\\
        DR200B5   &$0.10\pm 0.004$ &$0.08\pm 0.005$\\
        DR200B5C  &$0.10\pm 0.01$  &$0.0051\pm 0.0001$\\
        \tableline
        \tableline
    \end{tabular}
    \tablenotetext{1}{The average value followed by the standard deviation, over
    simulation time $50-90 t_0$. }
\end{table}

\subsection{Kinetic Energy Spectrum\label{section:result:Pk}}
\begin{figure*}[htbp]
\centering
     \begin{center}

      \subfigure[MRI turbulence]{%
            \includegraphics[width=0.49\textwidth]{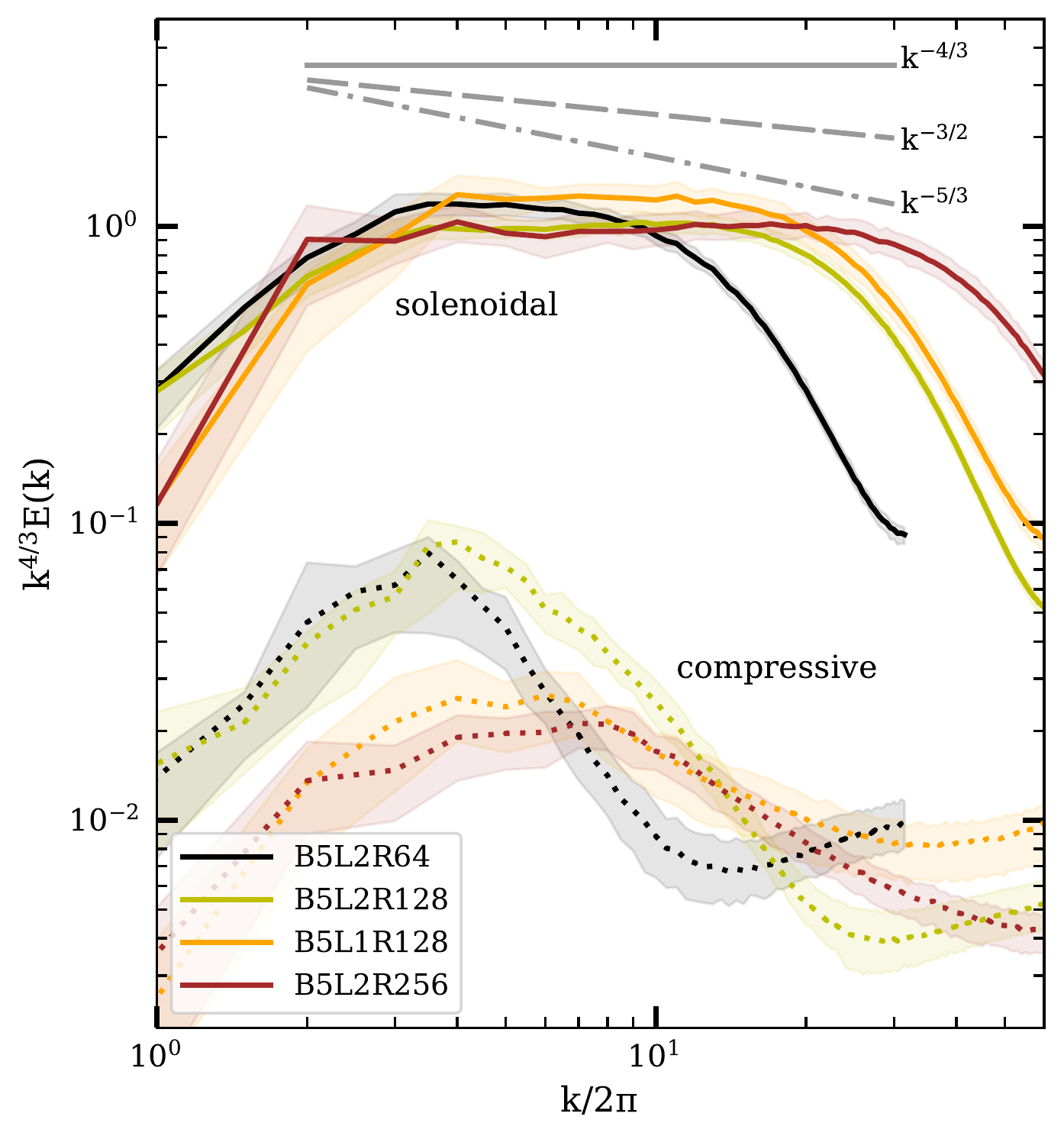}
        }
      \subfigure[isotropically driven turbulence]{%
           \includegraphics[width=0.49\textwidth]{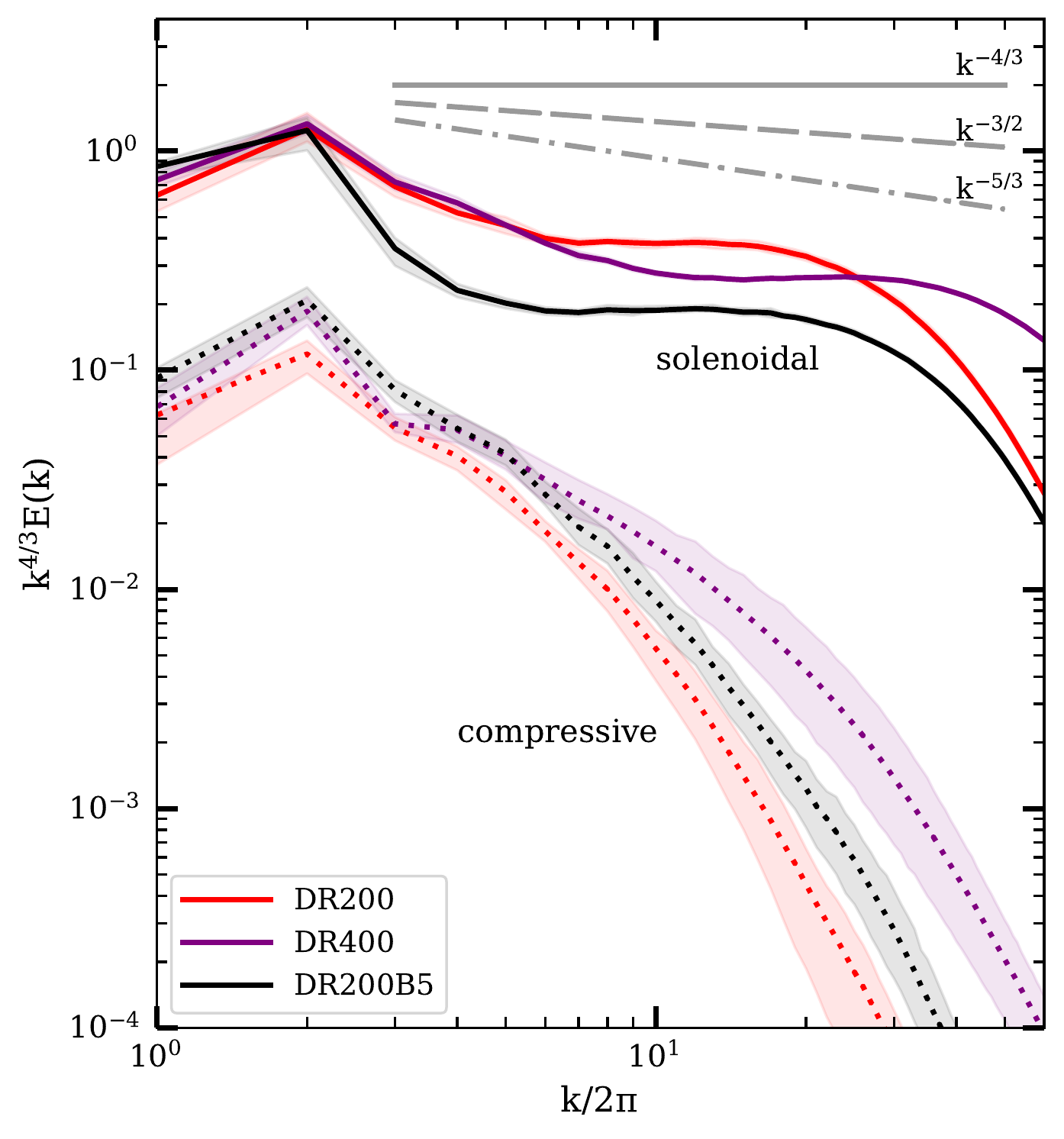}
        } 

    \end{center}
\caption{The kinetic energy spectrum 
    in the MRI (left panel) and isotropically driven (right panel)
    turbulence simulations. The $y$-axis is compensated by
    $k^{4/3}$, and normalized with the total turbulent kinetic energy 
    $\int \di k E(k)$ for each model.
    The time-averaged spectra in different simulation models
    are decomposed into solenoidal and compressive modes, shown with the solid and
    dashed lines (see figure legend).  The shaded
    area indicates the $1$-$\sigma$ dispersion among the different time snapshots. 
    The gray solid, dash, and dash-dotted lines show the power laws for
    $k^{-4/3}$, IK ($k^{-3/2}$), and Kolmogorov ($k^{-5/3}$) energy spectra.
    The total energy spectra in
    both the MRI and isotropically driven turbulence
    are dominated by the solenoidal modes, and show a power-law range 
    near the dissipation scale.
    Fitting of the turbulence energy spectra in the three high
    resolutions MRI simulations (B5L2R128, B5L1R128, B5L2R256)
    gives a slope of $E(k)\propto k^{-1.32\pm 0.02}$ in the power-law ranges,
    consistent with the $k^{-4/3}$ energy spectrum.}
\label{fig:Pk_MRI}
\end{figure*}

The turbulence energy spectra of the MRI simulations are shown in 
Figure \ref{fig:Pk_MRI} left panel. The solenoidal modes dominate,
containing more than
95\% of the total turbulent kinetic energy. This is not surprising, since the
linear analysis of the MRI already indicates that the instability can arise in
incompressible fluid \citep{BH1991}. The turbulence energy spectra $E(k)$
drops steeply at
$k/(2\pi) \gtrsim 0.1/\Delta x$ due to numerical dissipation, as no explicit
dissipation is included. $E(k)$ shows a power-law dependence on $k$ over the
middle range of $k$. Simulations with lower resolutions 
$L_0/\Delta x \leq 64$, such as B5L2R64, suffers from insufficient dynamical range,
and do not show a converged power-law slope due to numerical dissipation.
For higher resolution simulations showing a converged power-law slope, 
we fit simulations B5L2R128 and B5L1R128 over
$4<k/(2\pi)<10$ and simulation B5L1R256 over $4<k/(2\pi)<20$ with a power-law
function $E(k) \propto k^{-p}$ using the least square method, 
which yields an average of $p=1.32$. We estimate an error of $0.02$ 
for $p$, based on the error of fitting and the variation of $p$ when 
changing the range of $k$ for fitting. This slope is consistent with
$k^{-4/3}$, and is shallower than predictions from both the Kolmogorov ($k^{-5/3}$)
and the IK ($k^{-3/2}$) theories. 

In order to understand the $k^{-4/3}$ slope, we plot the energy spectrum from
driven turbulence simulations in Figure \ref{fig:Pk_MRI} right panel. With the
isotropic driving, the energy spectrum shows a slope consistent
with $k^{-4/3}$ in both the HD and MHD simulations. The slope is converged with
a resolution $L_0/\Delta x \geq 200$ in the HD simulations. This implies that
the $k^{-4/3}$ slope is an intrinsic property of the turbulence even in the
hydrodynamic case, and not a phenomenon that is only caused by magnetic fields.
With compressive driving, on the other hand, the energy spectrum
shows a steeper slope close to $k^{-7/3}$. This steepening of the energy
spectrum in compressive driving turbulence is also observed in numerical
simulations with a much higher resolution by \citet{Grete2018}.


There have been a great deal of work on the 
MHD turbulence driven by different mechanisms and with various numerical techniques.
The findings in the literature is summarized in Table \ref{table:literature}.
In all of the cases, $p$ smaller than $5/3$ is observed,
while a slope similar to $p=4/3$ is reported in many studies with different
turbulence driving mechanisms, equations of state, magnetic field strengths,
and numerical codes. 

The shallower energy spectrum near the dissipation scale has long been observed
by the fluid dynamics community in both numerical simulations and physical
experiment, albeit it is less known to the astrophysics community \citep[e.g.
reviews by][]{Sreenivasan1995, Alexakis2018}.
This is often referred to as the ``bottleneck'' effect, and is suggested to be 
caused by the helicity cascade \citep{Kurien2004}: the $k^{-4/3}$ slope can
arise when the helicity cascade timescale dominates over the energy cascade
timescale, and the bottleneck region is less pronounced when the turbulence
driving contains less helicity.
Recent high resolution simulations of hydrodynamic
turbulence by \citet{Ishihara2016} shows that the bottleneck region and a subsequent
``tilt'' extends over 2 -- 3 dex near the dissipation scale, before 
the Kolmogorov spectrum is observed on larger scales.

This cautions that a power-law region with converged slope, 
as is observed in our simulations,
does not mean that it is the inertial range. In the future, simulations with
at least an order of magnitude higher resolution are needed to explore the
inertial range of the MRI turbulence. This poses a significant challenge on
computational resources: the computation time step scales as $N^4$ ($N^3$ from
the spacial resolution in three dimensions, and $N$ from the reduced time
step) in grid-based codes, where $N$ is the number of resolution elements per
dimension. The spectral method is widely used for driven turbulence simulations,
but requires complex treatment to handle the background shear and boundary conditions
in shearing box simulations. \citet{LL2011} and \citet{Walker2016}
used the spectral code Snoopy to
simulate MRI turbulence, and obtained a highest resolution per $H$
about twice of that in this work.

\begin{table*}[htbp]
    \centering
    \caption{Kinetic and magnetic power spectra in MHD
    turbulence reported in the literature\tablenotemark{a}}
    \label{table:literature}
    \begin{tabular}{lcc ccc ccc}
        \tableline
        \tableline
        Ref. & Turbulence Generation &$\beta_\mr{init}$
        &Driven? &EOS &Codes &$N$
        &$p$ &$p_M$\\
        \tableline
        1\tablenotemark{b} &Kelvin-Helmholtz instability &$5000$
        &No &adiabatic &Athena &512
        &$1.33\pm 0.02$ &$1.62 \pm 0.02$\\
        2\tablenotemark{c} &shear-Alfven waves at large scales
        &$0.33$ &No &isothermal &PLUTO, VPIC &1152
        &$\sim 1.3$ &$\sim 1.3$\\
        3 &kinetic energy injection at large scales &$0.02$
        &Yes &adiabatic &Athena &1024
        &$1.38$ &$1.22$ \\
        4\tablenotemark{c} &external force at large scales
        &$5$, $72$ &Yes &isothermal &Enzo, Athena &1024
        &$\sim 4/3$ &$\sim 1.7$\\
        5\tablenotemark{c} &MRI with viscosity and resistivity &$400$
        &MRI &isothermal &ZEUS &512
        &$\sim 1.5$ &$\sim 0$ \\ 
        6\tablenotemark{c} &MRI with viscosity and resistivity &$1000$
        &MRI &incompressible &Snoopy &192
        &$\sim 1.5$ &- \\
        7\tablenotemark{c} &MRI with viscosity and resistivity &$1100$
        &MRI &incompressible &Snoopy &512
        &$\sim 1.5$ &$\sim 2$ \\
        \tableline
        \tableline
    \end{tabular}
    \tablenotetext{1}{Ref.: Literature references, (1)\citet{Salvesen2014},
    (2)\citet{Makwana2015}, (3)\citet{LS2009}, (4)\citet{Grete2017},
    (5)\citet{Fromang2010}, (6)\citet{LL2011}. (7)\citet{Walker2016}$.
    \beta_\mr{init}$: the plasma $\beta$ in the initial condition. ``Driven?'': 
    ``Yes'' means the turbulence is driven by manually injecting
    energy throughout the simulation time; ``No'' means that the turbulence
    develops from the initial condition and then decays without energy
    injection; and ``MRI'' means the turbulence is sustained by MRI.
    EOS: equation of state in the simulations. 
    Codes: Athena, ZEUS, Enzo and PLUTO are MHD grid-based codes,
    Snoopy is a MHD spectral code, and VPIC is a particle-in-cell code.
    $N$: Resolution of the simulation. If the resolution varies
    in different dimensions, the lowest resolution in all three dimensions is shown.
    $p$ and $p_M$: spectral indexes for kinetic and magnetic power
    spectra, $E(k)\propto k^{-p}$ and $E_M(k)\propto k^{-p_M}$.}
    \tablenotetext{2}{$p$ and $p_M$ are measured at the end of the
    simulation time. Throughout the time when turbulence decays until the
    simulations ends, the kinetic power spectrum index $p$ remains
    roughly constant, and the magnetic power spectrum index $p_M$ increases.}
    \tablenotetext{3}{The indexes of the power spectra are estimated
    without fitting.}
\end{table*}

\begin{figure*}[htbp]
\centering
\includegraphics[width=\linewidth]{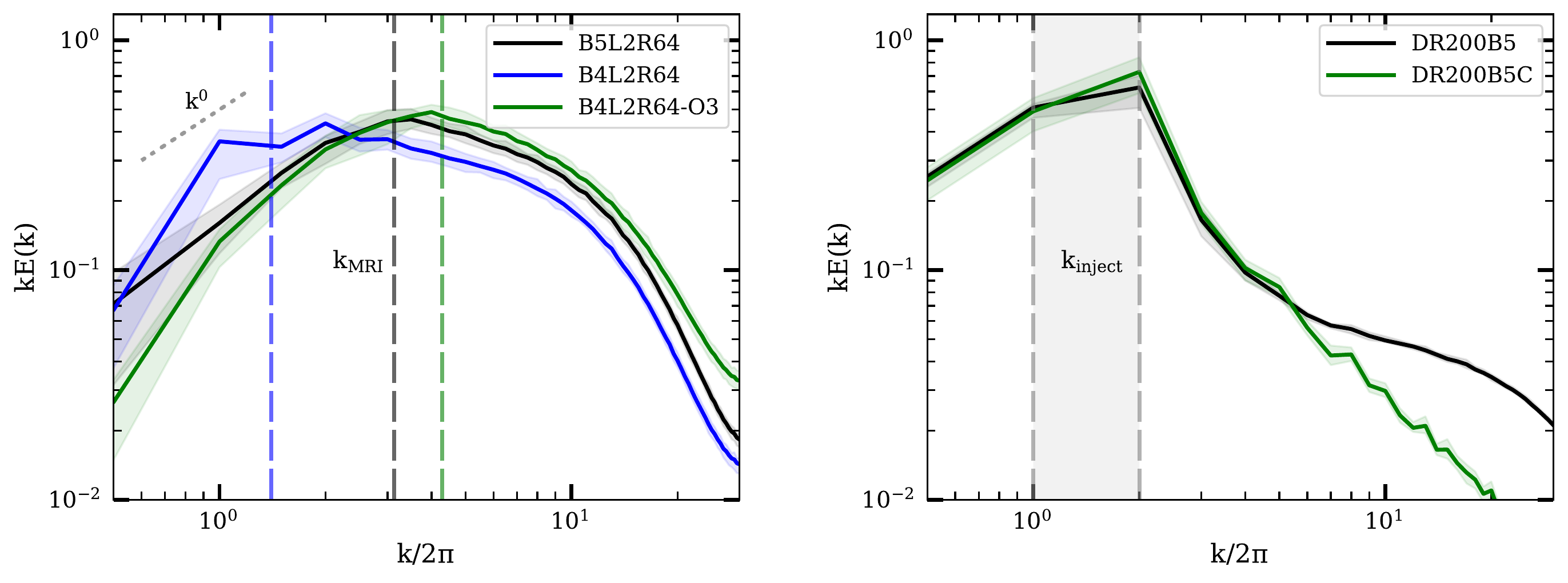}
    \caption{Kinetic energy spectra compensated by $k$,
    for MRI (left panel) and driven turbulence (right panel) simulations.
    The $y$-axis is normalized with the total turbulent kinetic energy 
    $\int \di k E(k)$ for each model. The solid lines show 
    the time-averaged energy
    spectrum in different simulation models (see figure legend), and the shaded
    area indicates the $1$-$\sigma$ dispersion among different time snapshots. The
    vertical dashed lines in the left panel denote the estimation 
    for the fastest growing mode
    for the MRI, $k_\mathrm{MRI}$ (Equation (\ref{eq:kMRI})). In the right
    panel, the vertical dashed lines bounding the shaded area indicate
    the range of scales where kinetic energy is injected to drive the
    turbulence. The peak of
    $kE(k)$, representing the maximum kinetic energy density,
    matches the expected energy injection scale of
    $k_\mathrm{MRI}$ and $k_\mathrm{inject}$.}
\label{fig:Pk_injection}
\end{figure*}

Another important parameter for
the kinetic energy spectrum is the injection scale. \citetalias{OC2007} assume
that the injection scale $k_L$ is the disk scale-height $H$, and $E(k)=0$ for
$k<k_L$. In reality, $E(k)$ never drops to zero in MRI turbulence, as shown in
Figure \ref{fig:Pk_MRI}. Therefore, one can define the effective injection scale as
the peak of $kE(k)$ in Figure \ref{fig:Pk_injection}, 
the scale where most of the kinetic energy is concentrated at. 
We found that the injection scale in MRI
turbulence can be approximated by an estimation of the fastest growing mode of
the linear MRI \citep{BH1991},
\begin{equation}\label{eq:kMRI}
    k_\mathrm{MRI} = \sqrt{\frac{15}{16}}\frac{\Omega}{\langle v_{A,z}
    \rangle}.
\end{equation}
$k_\mathrm{MRI}$ for simulations with different initial magnetic
field strengths and orbital frequencies are shown as the vertical dashed lines
in the left panel of Figure \ref{fig:Pk_injection}. In the right panel of
Figure \ref{fig:Pk_injection}, we show ath the injection scale in the driven
turbulence simulations is, indeed, close to the peak of $kE(k)$.  
We note that contrary to the assumption in \citetalias{OC2007} that the
injection scale is at the disk scale height $H$, $k_\mathrm{MRI}$ is determined
by the magnetic field strength instead of $H$.  For a weak vertical 
magnetic field in our simulations, $\langle v_{A,z} \rangle$ is 
smaller than the sound speed, and the length scale $1/k_\mathrm{MRI}$ 
is smaller than $H$.

\subsection{Magnetic Energy Spectrum\label{section:result:Pk}}
\begin{figure*}[htbp]
\centering
\includegraphics[width=\linewidth]{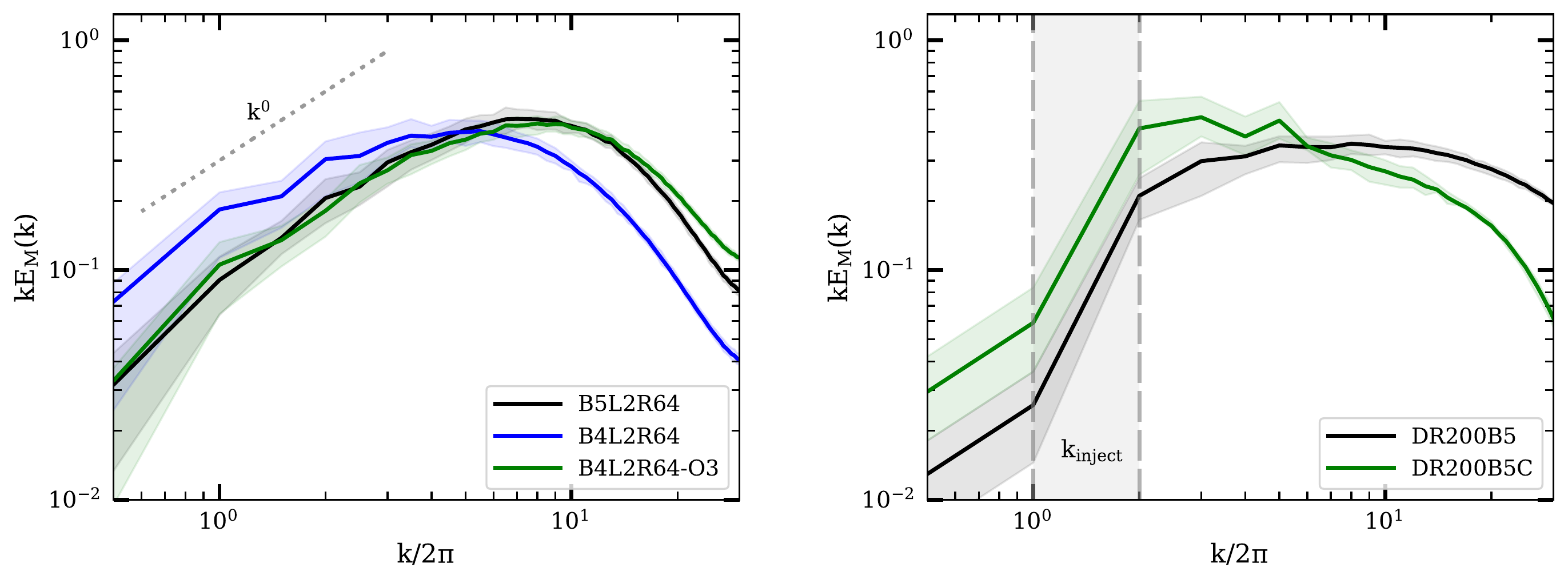}
    \caption{The magnetic energy spectrum $E_M(k)$ compensated by $k$, for MRI (left
    panel) and driven turbulence (right panel) simulations, similar to the
    kinetic energy spectrum in Figure \ref{fig:Pk_injection}. The peak of the
    magnetic energy $kE_M(k)$ is located near the dissipation scale for MRI
    simulations (left panel), and near the injection scale for driven
    turbulence simulations (right panel). The dotted line in the left panel
    shows that $E_M(k)$ in the MRI turbulence is close to a constant in a wide
    range of $k$ before the numerical dissipation becomes important.
    }
\label{fig:EM}
\end{figure*}

Unlike the kinetic energy spectrum, the magnetic energy spectrum does not show
any power-law behavior (Figure \ref{fig:EM}). $E_M(k)$ is roughly constant
across a wide range of $k$, until it drops off quickly at larger $k$ due to
numerical dissipation. In MRI turbulence, the magnetic energy $kE_M(k)$ peaks
near the dissipation scale (Figure \ref{fig:EM}, left panel),
possibly due to the injection of magnetic energy at small scales by the dynamo
mechanism proposed by \citet{Schekochihin2002}. Similar results also have been 
found in MRI simulations by \citet{Fromang2010} with explicit resistivity
and viscosity. There is also evidence that the position of the peak 
may depend on the magnetic Prandtl number, and not
always be at the resistive scale \citep{Subramanian1999, Haugen2003}. In ideal
MHD simulations, it is difficult to quantify the effective Prandtl number from
numerical dissipation,
and therefore future simulations with explicit dissipation are
needed to address this issue.

\begin{figure*}[htbp]
\centering
     \begin{center}
      \subfigure[MRI]{%
            \includegraphics[width=0.48\textwidth]{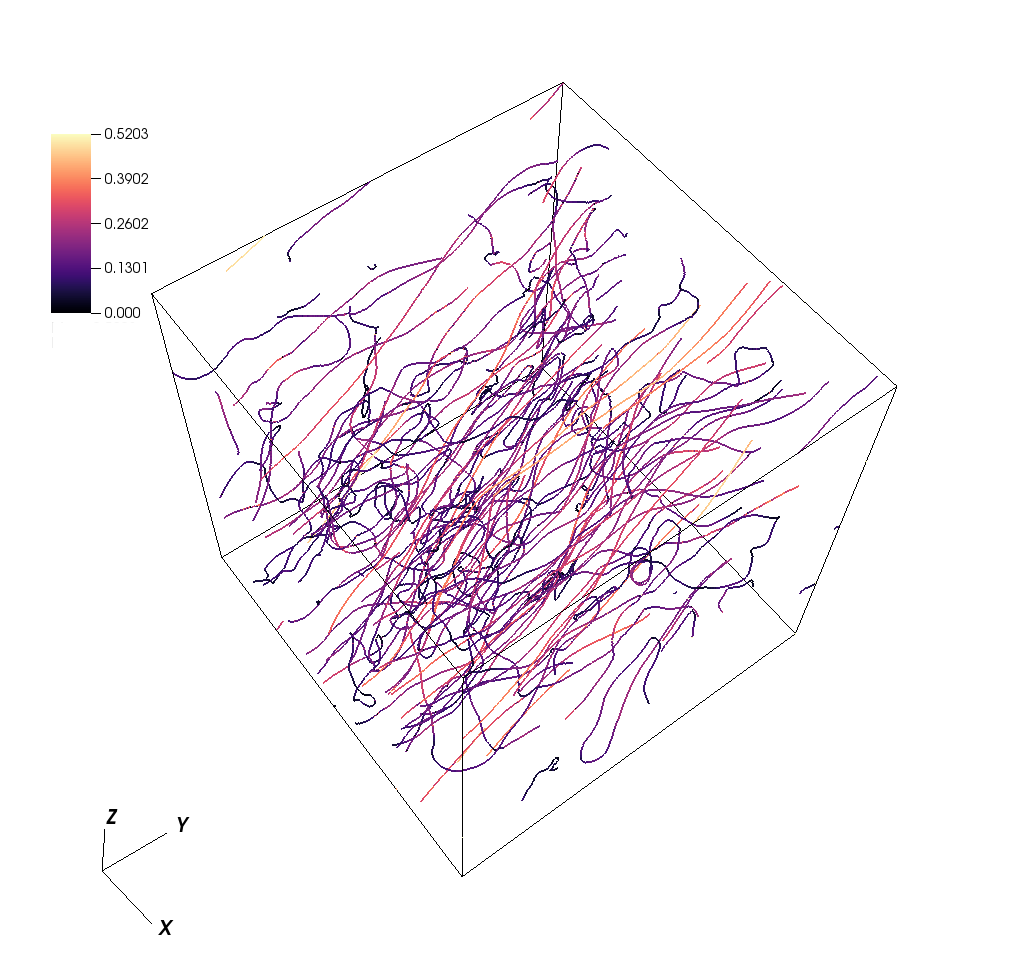}
        }
      \subfigure[driven turbulence]{%
           \includegraphics[width=0.48\textwidth]{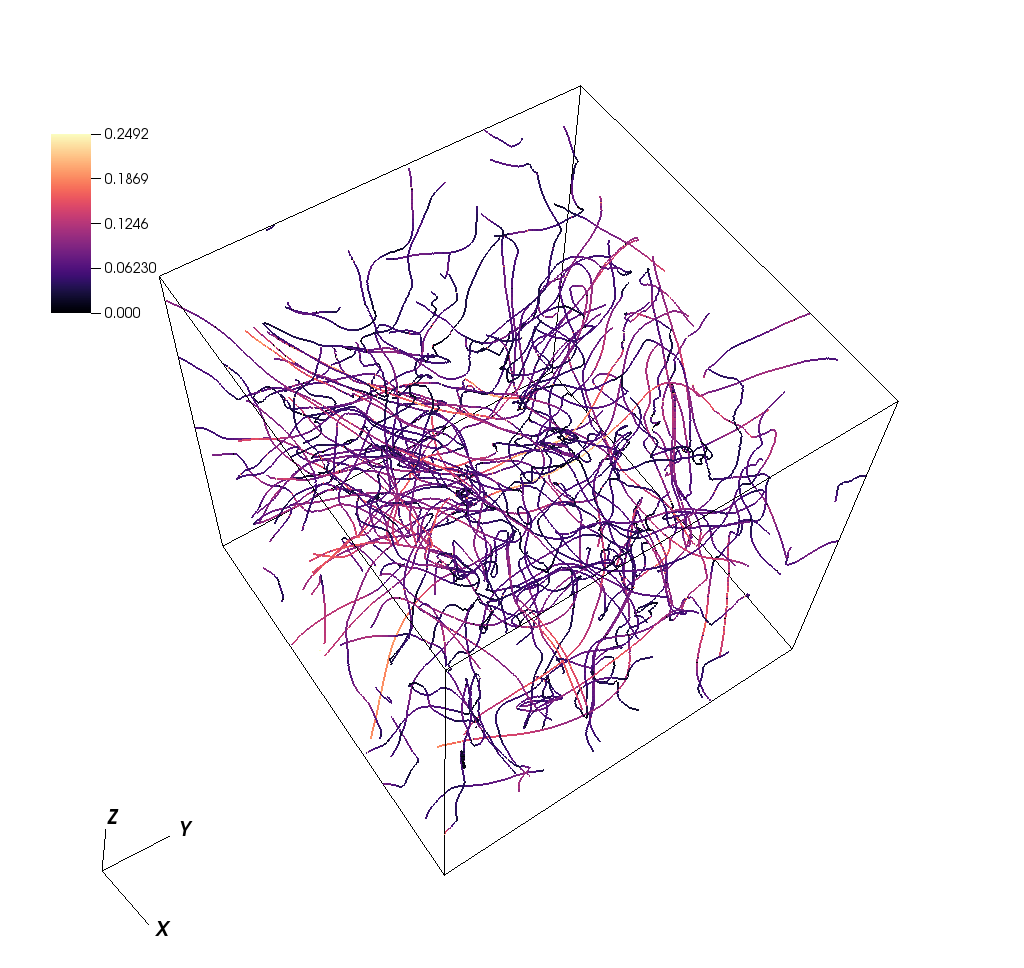}
        } 

    \end{center}
    \caption{Visualization of the magnetic field lines in (a) MRI simulation
    B5L1R256 and (b) driven turbulence simulation DR200B5. The snapshots are
    taken at $t=90t_0$, when both simulations already reached steady state.
    The simulation box-size is $L_0$ on all sides. The color of the streamline
    shows the magnetic field strength (stronger field in lighter colors, see
    legends).  Compared to the driven turbulence, the magnetic field in the MRI
    turbulence has more power on the small scales (Figure \ref{fig:EM}),
    and a mean field along y-axis due to the shearing motion.
        \label{fig:Bfl}}
\end{figure*}

In driven turbulence
simulations, $kE_M(k)$ peaks at large scales, where perturbations are driven
(Figure \ref{fig:EM} right panel), consistent with injection of magnetic energy
by the perturbation. Figure \ref{fig:Bfl} illustrates the magnetic field lines,
which shows visually that small-scale perturbations of the magnetic field are more
prominent in the MRI turbulence than in the driven turbulence.

\subsection{The Effect of Stratification}
In order to investigate the effect of stratification on the turbulence, we
carry out stratified simulation B5L2R32\_S, with the initial conditions in the
mid-plane similar to those the unstratified simulation B5L2R32. Due to the larger
box and fast outflow at the $z$-boundary, the stratified simulation is much
more computationally costly, and thus the resolution is limited.
However, this still allows us to compare the positions of the peaks 
of the energy spectra in the stratified and unstratified models.

\begin{figure*}[htbp]
\centering
\includegraphics[width=\linewidth]{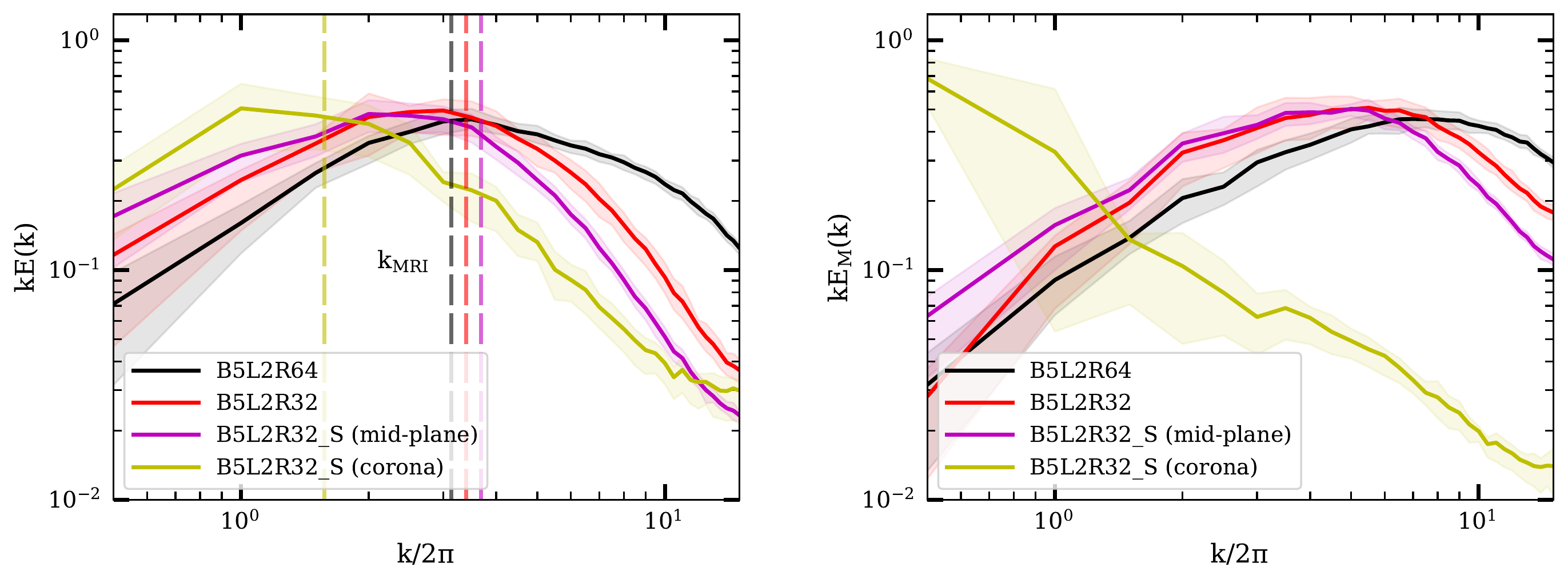}
    \caption{Kinetic ({\sl left} panel) and magnetic ({\sl right} panel)
    energy spectra
    compensated by $k$, for stratified and unstratified simulations (see
    legend), similar to the left panels in Figures \ref{fig:Pk_injection} and
    \ref{fig:EM}. The energy spectra are shown in the mid-plane ($-1<z<1$) and
    upper corona ($1<z<3$, the spectra for $-3<z<-1$ are
    very similar) for the stratified simulation B5L2R32\_S. The shapes of the
    spectra are similar between the stratified simulation B5L2R32\_S in the
    mid-plane region, and the unstratified simulation B5L2R32. The fastest growing
    mode for the MRI, $k_\mathrm{MRI}$ (Equation (\ref{eq:kMRI})), reasonably 
    matches the maximum of the kinetic energy density in both the mid-plane and the
    corona regions. The magnetic energy spectrum in the corona region, however,
    is distinctively different from that in the disk mid-plane, and peaks at
    much larger scales.}
\label{fig:Ek_strat}
\end{figure*}

Figure \ref{fig:Ek_strat} shows the kinetic and magnetic energy spectra in the
stratified simulation B5L2R32\_S and unstratified simulation B5L2R32. The
higher-resolution unstratified simulation B5L2R64 is also plotted for
reference. Both the kinetic and magnetic energy spectra are similar between
B5L2R32\_S in the midplane and the unstratified
simulation B5L2R32, justifying the use of the unstratified simulations 
for the mid-plane conditions. The maximum of the kinetic energy density
is similar to $k_\mathrm{MRI}$ in both the disk mid-plane and corona,
although the gas density drops by nearly an order of magnitude from the disk
mid-plane to the corona region. The magnetic energy spectrum in the corona
region, however, differs significantly from the disk mid-plane, and peaks
instead at the largest scales. This is similar to the findings by \citet{Nauman2014}
and \citet{Blackman2015}, who pointed out the potential importance of non-local
transport in the disk. Thorough investigations of the impact of stratification
and global geometry on the shape of the energy spectrum, however, require
global disk simulations with very high resolution, which is beyond the scope of
this paper.

\subsection{Auto-correlation Time\label{section:results:tauk}}
\begin{figure*}[htbp]
\centering
\includegraphics[width=\linewidth]{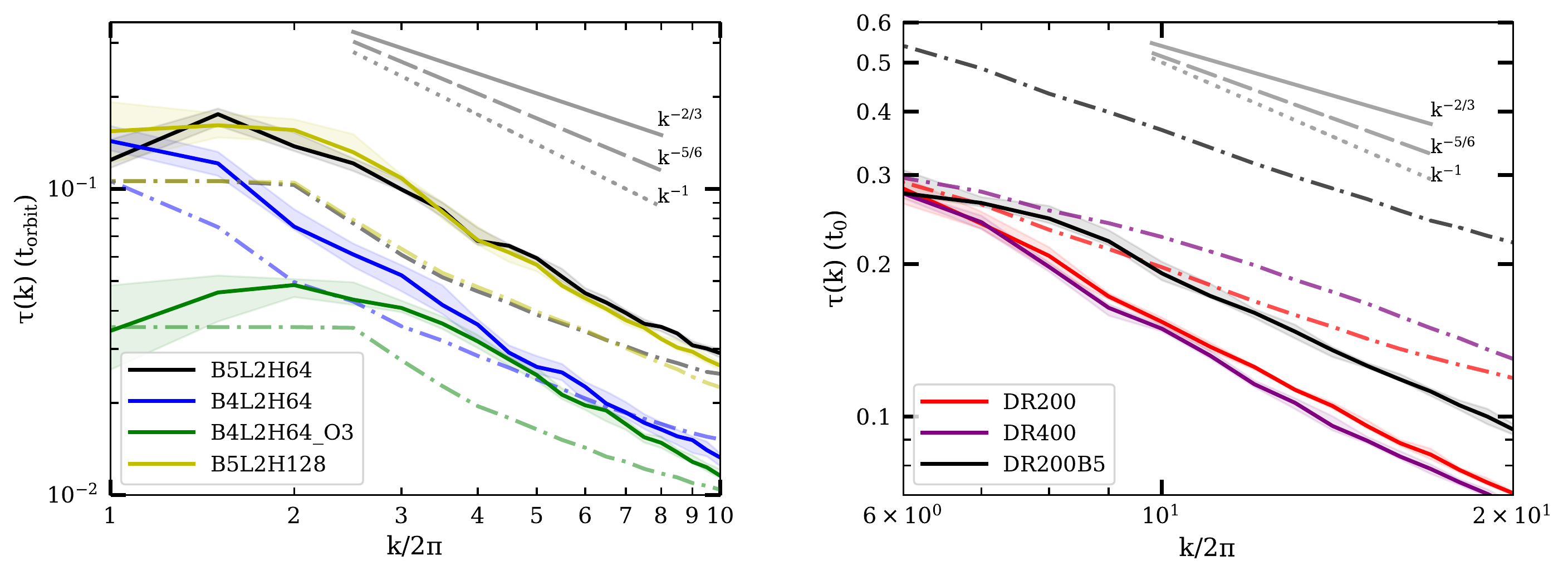}
    \caption{The eddy auto-correlation time obtained from the MRI (left panel)
    and driven turbulence (right panel) simulations.
    The solid lines show the time-averaged $\tau(k)$ in
    different simulation models (see figure legend), and the shaded
    area indicates the $1$-$\sigma$ dispersion among different time snapshots.
    The dashed-dotted lines of the corresponding color 
    shows the expected $\tau(k)$ in each model: the minimum of 
    the shearing timescale and the eddy crossing time in the MRI
    simulations (left panel), and the eddy crossing time only in the
    driven turbulence simulations (right panel). 
    The gray solid, dashed and dotted lines indicate a power-law slope of
    $k^{-2/3}$ (Kolmogorov eddy crossing time), $k^{-5/6}$ (eddy crossing time
    for $k^{-4/3}$ energy spectrum ), and $k^{-1}$. 
    }
\label{fig:tauk}
\end{figure*}

The auto-correlation of turbulence eddies as a function of time
can be well fitted by Equation (\ref{eq:tauk}) in the
power-law range of the kinetic energy spectrum (Figures \ref{fig:Pk_MRI}),
where the influence from energy injection at large scales and
numerical dissipation at small scales is small. Figure \ref{fig:tauk} shows a
summary of $\tau(k)$ obtained from the simulations. $\tau(k)$ in the MRI
turbulence is roughly a constant at large scales,
limited by the shearing timescale $\tau_\mathrm{shear}=1/\Omega$, 
and shows a power-law drop off at small scales, limited by the eddy crossing
time $\tau_\mathrm{cross} = 1/(k\delta v(k)) = 1/(k\sqrt{2k E(k)})$.
The power-law slope is converged for the MRI simulation B5L2H64 and B5L2H128,
as well as the driven turbulence simulation DR200 and DR400.
Fitting with $\tau(k)\propto k^{-m}$ in the power-law range gives a slope of
$m=1.11\pm 0.07$ for the MRI simulation B5L2H64 and B5L2H128 between $3<k/(2\pi)<9$,
and $m=1.16\pm 0.02$
for the driven turbulence simulation DR200 and DR400 between $9<k/(2\pi)<20$.
This is somewhat steeper than the slope of $\tau(k)\propto k^{-2/3}$
predicted from the eddy crossing time for a Kolmogorov spectrum and of
$\tau(k)\propto k^{-5/6}$ for a
$E(k)\propto k^{4/3}$ spectrum (as found in Section \ref{section:result:Pk}).

\section{Conclusions}\label{section:conclusions}
In this paper, we investigate the key underlying assumptions for the turbulence
properties governing the grain growth in protoplanetary disks. We
carry out ideal MHD shearing-box simulations to investigate the MRI turbulence in 
protoplanetary disks. We also perform HD and MHD driven turbulence simulations
to compare with the MRI turbulence.

Both the energy spectrum $E(k)$ and the auto-correlation time $\tau(k)$
of the MRI turbulence in our simulations deviate from those in 
the Kolmogorov turbulence, which is widely adopted by the current grain collisional
velocity calculations such as \citetalias{OC2007}.
The main findings of this paper are summarized as follows:
\begin{enumerate}
    \item The MRI turbulence is largely solenoidal.
        We observe $E(k) \propto k^{-4/3}$ in the MRI
        turbulence, as well as in an isotropically
        driven turbulence with and without magnetic
        field, over more than 1 dex near the dissipation scale (Figures
        \ref{fig:Pk_MRI}). This power-law slope appears
        to be converged in terms of numerical
        resolution, and to be due to the bottleneck effect.
    \item The kinetic turbulence energy spectrum 
        $E(k)$ converges at a much higher resolution ($128$ cells per $H$)
        than the turbulence $\alpha$-parameter ($32$ cells per $H$). 
    \item The kinetic energy in the MRI turbulence peaks 
        at the fastest growing mode of the MRI (Figure \ref{fig:Pk_injection}),
        which does not necessarily match the disk scale height as assumed by
        \citetalias{OC2007}.
        In contrast, the magnetic energy peaks at the dissipation scale (Figure
        \ref{fig:EM}).
    \item The magnetic energy spectrum $E_M(k)$ in the MRI turbulence 
        does not show a clear power-law range, and is almost constant over
        approximately 1 dex near the dissipation scale (Figure \ref{fig:EM}).
    \item The turbulence autocorrelation time $\tau(k)$ in the MRI turbulence
        is nearly constant at large scales,
        limited by the shearing timescale, and shows a power-law drop 
        close to $k^{-1}$ at
        small scales, with a slope slightly steeper than the eddy crossing time
        (Figure \ref{fig:tauk}).
\end{enumerate}

Due to the bottleneck effect near the dissipation scale, the power-law
slopes of $E(k)$ and $\tau(k)$ observed in our simulations may
not represent those in the inertial range of the turbulence. Still, only slight
changes in the shapes of $E(k)$ and $\tau(k)$ can have a 
significant effect on the grain collisional velocities. 
In a future paper, we plan to investigate the potential impact of 
the turbulence properties on the grain size evolution in protoplanetary
disks.

\section{Acknowledgement}
We thank the anonymous referee for a constructive review,
which helped to improve the quality and clarity of this paper.
M. Gong thanks Xuening Bai and Jake Simon for their generous help and advices 
on the use of the {\sl Athena} code and related scientific discussions,
Geoffroy Lesur for his helpful suggestions on turbulence analysis,
and Chang-Goo Kim for his
implementation of the turbulence driving module in the {\sl Athena++} code.

\appendix
\section{Calculation of Turbulence Properties from Numerical
Simulations\label{section:method:turb}}
The turbulence power
spectrum $P(k)$ and auto-correlation time $\tau(k)$ determine the collisional
velocities between dust grains. In \citetalias{OC2007}, $P(k)$ and $\tau(k)$ are
assumed to follow the analytic expression of Kolmogorov turbulence:
$E(k)\equiv 4\pi k^2 P(k)\propto k^{-5/3}$ and $\tau(k)=1/(k \sqrt{2kE(k)})$.
To study the MRI turbulence in protoplanetary disks,
we derive $P(k)$ and $\tau(k)$ from numerical simulations,
using the procedures described below.

As the simulations are preformed on a discrete grid, we use the discrete fast
Fourier transform to calculate $\widetilde{y}(\mathbf{k})$ for the physical
variables $y(\mathbf{x})$ in our
simulations. The resolution of $\widetilde{y}(\mathbf{k})$ in $\mathbf{k}$ space is
$\di k_i = 2\pi/L_i,~i=x, y, z$.  $\widetilde{y}(k)$ is then calculated by averaging
over co-centric shells of width $\di k=\mathrm{max}\{\di k_x, \di k_y, \di k_z\}$ 
around the origin in $\mathbf{k}$ space . The kinetic
power spectrum $P(k)$ is then obtained from
Equation (\ref{eq:Pk}). To reduce noise, we average $P(k)$ over a period of
time when the simulations already reach quasi-equilibrium 
(see Section \ref{section:results:MRI}).
We produce outputs of the simulations at intervals of $t_\mr{orbit}$, and the average
$P(k)$ is calculated from the outputs during simulation time $40t_\mr{orbit}-80t_\mr{orbit}$ for model
B5L2R32-s and $90t_\mr{orbit}-130t_\mr{orbit}$ for all other models. 

$\tau(k)$ is obtained by fitting Equation (\ref{eq:tauk}), 
where the values of $f(\Delta t)$
are calculated directly from the numerical simulations, 
and $|\widetilde{\delta \mathbf{v}}(\mathbf{k},t_s)|^2$ 
and $\tau(k)$ are treated as constant parameters obtained from the 
least-square fitting of $f(\Delta t)$ as a function of $\Delta t$. We output
simulations and calculate $f(\Delta t)$ 
at small time intervals $\di t_\mathrm{out} \ll \tau(k)$, to obtain
enough time resolution in $\Delta t$ to accurately fit $\tau(k)$. We
also sample a range of $t_s$ to characterize the noise in $\tau(k)$.
We transfer the turbulent velocity field $\delta \mathbf{v}$ onto the
shearing frame $y'=y + q\Omega x \Delta t$, so $\tau(k)$ is only determined by the
turbulent motions.

\bibliographystyle{apj}
\bibliography{apj-jour,/Users/munangong/myLatex/all}
\end{document}